\documentclass[twocolumn,prl,showpacs,nofootinbib]{revtex4-1}
\usepackage{graphicx}
\usepackage[utf8x]{inputenc}
\usepackage{dcolumn}
\usepackage{bm}

\newcommand{\beq}{\begin{equation}}
\newcommand{\eeq}{\end{equation}}
\newcommand{\beqa}{\begin{eqnarray}}
\newcommand{\eeqa}{\end{eqnarray}}

\begin{document}

\noindent\textbf{Are GRB 090423 and Similar Bursts due to Superconducting Cosmic Strings?}\\[-1ex]

The superconducting cosmic string (SCS) is one of the earliest suggested cosmological sources for the violent gamma-ray bursts (GRBs) \cite{BPS87}. An interesting feature of the SCS model is the rapid increase of the GRB rate with the redshift. Indeed in a recent paper by Cheng et al. \cite{Cheng10}, it was shown that the rate of GRBs at redshifts $z>6$, significantly deviating from the star formation rate, is nicely consistent with the SCS model, implying that these high redshift GRBs may be mainly produced by the SCSs moving relativistically in the Universe. The physical parameters of the SCS are then constrained by the duration, the luminosity and the rate of the high redshift events such as GRB 080913 and GRB 090423 \cite{Cheng10}. {\it There is, however, a key constraint that has been ignored.} In the SCS scenario the electromagnetic blast wave driven by the cusp is concentrated in a narrow cone with an opening angle $ \sim 1/\gamma_{\rm l.o.s.}$ \cite{BPS87,Cheng10}, where $\gamma_{\rm l.o.s.} \sim 10^{3}$ is the bulk Lorentz factor of segment of the cusp along the line of sight.
The opening angle of the prompt $\gamma-$ray emission region is taken to be the same (i.e., $\sim 1/\gamma_{\rm l.o.s.}$), as widely adopted in the estimate of the luminosity and the rate of the SCS GRBs (for instance see Eq.(1) and Eq.(5) of \cite{Cheng10}). {\it However the typical opening angle of GRBs} ($\theta_{\rm GRB} \sim 0.1$) {\it is found to be two or more orders of magnitude wider.}  For example, for GRB 090423 at $z=8.2$, an opening angle $\sim 0.2$ is needed to reproduce the plentiful X-ray/optical/radio afterglow data \cite{Chandra10}. Such a huge inconsistence imposes a tight constraint on the SCS GRB models.

Cheng et al.\cite{ChengReply} further suggested that the outflow driven by SCSs is very wide and the energy distribution is given by
\begin{equation}
\epsilon(\theta)=
\left\{
\begin{array}{ll}
{\epsilon_{0}~  \theta _{\rm m}^{3} \theta ^{-3}}, & \theta >\theta _{\rm m}, \\
{\epsilon _{0} }, & \theta \leq \theta _{\rm m},
\end{array}
\right.
\label{Eq:Cheng}
\end{equation}
where $\epsilon _{0}\sim 5\times 10^{52}~{\rm erg}~(\theta_{\rm l.o.s.}/\theta_{\rm m})^{3}$, the angle between the symmetric axis of the ejecta and the line of sight $\theta_{\rm l.o.s.} \sim 1/\gamma_{\rm l.o.s.} \sim 10^{-3}$ and $\theta_{\rm m}\sim 10^{-8}$ \cite{ChengReply}. These authors then speculated that such a highly structured jet might be able to reproduce the afterglow emission of GRB 090423. Below we show it seems not the case.

The highly structured jet described by Eq.(\ref{Eq:Cheng}) has a very energetic core as narrow as $\sim 10\theta_{\rm m} \sim 10^{-7}$, within which the ejecta energy is about $93\%$ of the total ($\sim 10^{52}$ erg). Its initial Lorentz factor is extremely high since in the SCS scenario there is no baryon loading. The radiation of such an ``off-beam" narrow energetic core will not be observed until it gets decelerated to a bulk Lorentz factor $\gamma \sim 1/\theta_{\rm l.o.s.}$. The corresponding time can be estimated to be $t\sim 8~{\rm day}~ f_1(\epsilon_{0,67.7}/n)^{1/3}\theta_{\rm l.o.s.,-3}^{8/3}$ if the sideways expansion is unimportant \cite{Wei03}, where $f\equiv1+z$, $n$ is the number density of circumburst medium and the convention $Q_{\rm x} = Q/ 10^{\rm x}$ has been adopted. The emergence of the forward shock emission of such an energetic core will produce an unambiguous rebrightening, as was widely found in previous studies \cite[e.g.,][]{Wei03,Kumar03,Piran04}.
 The lack of any signature of rebrightening in the optical and X-ray afterglow lightcurves of GRB 090423 disfavors the highly structured jet model.

If the sideways expansion velocity of the decelerating ejecta is comparable to that of the sound speed of the shocked medium, the dynamics will be modified. The bulk Lorentz factor is given by
\begin{equation}
\gamma  \sim 300(\epsilon _{0,67.7}/n)^{1/6}(t_{\rm day,-1}/f_1)^{-1/2}\theta _{\rm m,-8}^{1/3},
\end{equation}
where $t_{\rm day}$ is the time in units of day. In this case, the emission of the sideways-expanding energetic core is observable at a time $t\gtrsim  0.1$ day. Therefore no significant rebrightening is expected at late times, in agreement with the data. With the X-ray spectrum $F_{\nu}\propto \nu^{-1.05\pm 0.1}$ at $t\sim 0.1$ day, the forward shock X-ray emission should be above both the cooling frequency and the typical synchrotron radiation frequency supposing the accelerated electrons have a power-law spectral index $\sim -2.1$. Following \cite{Piran04}, it is straightforward to show that the forward shock X-ray emission should drop with time as $t^{-2.1\pm 0.2}$, significantly steeper than the observed lightcurve.

So far we have shown the highly structured jet described by Eq.(\ref{Eq:Cheng}) is hard to reproduce the
afterglow data of GRB 090423. suggesting that the SCS GRB outflow
with a very energetic narrow core is not supported by
current observations. If the very high redshift bursts, such
as GRB 080913 and GRB 090423, are indeed produced by
superconducting cosmic strings, a prompt process, unclear
so far, that can significantly broaden the initial ejecta is
needed.

We acknowledge Professor Venyamin Berezinsky for discussion on the opening angle of the SCS GRBs.

\vspace{2ex}
\noindent Yu Wang, Yi-Zhong Fan,$^\ast$ and Da-Ming Wei\\
\small{\indent Key Lab of Dark Matter and Space Astronomy

 Purple Mountain Observatory

 Chinese Academy of Sciences
210008 Nanjing, P. R. China }\\[-1ex]

$^\ast$Corresponding author.

~~yzfan@pmo.ac.cn

\end{document}